\documentclass{aa}
\usepackage{txfonts}
\usepackage{natbib}
\usepackage{graphicx}
\begin{document}

\title{PKS 0537-286, carrying the information of the environment of SMBHs in the early Universe}

\author{{E. Bottacini\inst{1}, M. Ajello\inst{2}, J. Greiner\inst{1}, E. Pian\inst{3,5}, A. Rau\inst{1}}, E.
Palazzi\inst{4},  S. Covino\inst{6}, G. Ghisellini\inst{6}, T. Kr\"uhler\inst{1,7}, A. K\"upc\"u Yolda\c{s}\inst{5}, N. Cappelluti\inst{1}, P. Afonso\inst{1}}

\institute{
		 Max-Planck Institut f\"ur Extraterrestrische Physik, Giessenbachstrasse,  85741, Garching, Germany
     \and Stanford Linear Accelerator Center/KIPAC, 2572 Sand Hill Road, Menlo Park, CA 91125, USA
     \and Istituto Nazionale di Astrofisica, Via G. B. Tiepolo 11, 34143 Trieste, Italy
     \and Istituto Nazionale di Astrofisica, Via Gobetti 101, 40129 Bologna, Italy
     \and ESO, Karl-Schwarzschild-Strasse 2, 85748 Garching, Germany
     \and Istituto Nazionale di Astrofisica, Osservatorio Astronomico di Brera, via E. Bianchi 46, 23807 Merate, Italy
     \and Universe Cluster, Technische Universit\"{a}t M\"{u}nchen, Bolzamannstrasse 2, 85748 Garching, Germany
          }

\offprints{E. Bottacini,
\email{eub@mpe.mpg.de}}
\date{Received September 17, 2009 / Accepted September 25, 2009}

\authorrunning{E. Bottacini et al.}
\titlerunning{PKS 0537-286, a multifrequency campaign}

\abstract {The high-redshift (z = 3.1) blazar PKS 0537-286, belonging to the flat spectrum radio quasar blazar subclass, is one of the most luminous active galactic nuclei(AGN) in the Universe. Blazars are very suitable candidates for multiwavelength observations. Indeed, the relativistic beaming effect at work within the jet enhances their luminosity. This in turn allows the properties of the extragalactic jets, the powering central engine, and the surrounding environment to derived.}{%
Our aim is to present the results of a multifrequency campaign from the near-IR to hard X-ray energies on PKS 0537-286 and give insight into the physical environment where the radiation processes take place.}{%
We observed the source at different epochs from 2006 to 2008 with INTEGRAL and {\it Swift}, and nearly simultaneously with ground-based optical telescopes. We also analyzed two archival spectra taken with XMM-Newton in 1999 and 2005. A comparative analysis of the results is performed.}{%
The X-ray continuum of the blazar, as sampled by XMM, is described by a power law of index $\Gamma$ = 1.2, modified by variable absorption at the soft X-rays, as found in other high-redshift QSOs. Modest X-ray continuum variability is found in the \textit{Swift} observations. The combined {\it Swift}/BAT and {\it Swift}/XRT spectrum is very hard ($\Gamma$ = 1.3). This, together with the non simultaneous EGRET detection and the more recent non detection by Fermi-LAT, constrains the peak of the high-energy component robustly. The optical/UV data, heavily affected by intervening Ly $\alpha$ absorption, indicate the presence of a bright thermal accretion disk that decreased in luminosity between 2006 and 2008. We infer from this a reduction of the BLR radius. When taking this into account, the 2006 and 2008 SEDs are compatible with a model based on synchrotron radiation and external inverse Compton scattering where the accretion-disk luminosity decreases between the 2 epochs by a factor 2, while the bulk Lorentz factor remains unchanged and the magnetic field changed only marginally.}{}
\keywords{quasars: individual (PKS 0537-286) -- galaxies: active -- X-rays: galaxies -- galaxies: high-redshift -- radiation mechanisms: nonthermal}
\maketitle

\section{Introduction}
Simultaneous multiwavelength studies of blazars offer a tool for determining the properties of extragalactic jets, their powering central engine, and the surrounding environment. The alignment of the jet with respect to the observer's line of sight \citep{antonucci93,urry95} makes blazars very luminous from radio frequencies to gamma-ray energies \citep{ulrich97}. Indeed, the beaming in these relativistic outflows \citep{rees66} enhances the luminosity. While the jet propagates, part of its power is dissipated via cooling and acceleration processes of protons and leptons. The emitting particles, together with the relativistic beaming, give rise to the typical two broad hump structure of the spectral energy distribution (SED). The low-energy emission is understood and is believed to be produced through synchrotron radiation of relativistic electrons. In contrast, the nature of the high-energy hump is still being debated. A possible explanation is that the same electron population that is responsible for the low-energy hump generates the high-energy hump through the inverse-Compton mechanism \citep{sikora94, dermer93, ghisellini96}. Another model predicts that the seed photons for the inverse-Compton are external to the jet \citep{sikora94, dermer92}. Alternative explanations are proton-initiated cascades \citep{mannheim93} or proton-synchrotron emission from ultra-relativistic protons \citep{muecke03, aharonian00}.

High-redshift blazars are important tracers of the evolution of super massive black holes (SMBH) in the early Universe. \citet{boettcher02} link the two blazar subclasses, namely flat spectrum radio quasars (FSRQ) and BL the Lacertae (BL Lac) objects, which are the beamed counterparts of low- and high-luminosity radio galaxies, respectively \citep{wall97, willott01}, through the gradual depletion of the circum-nuclear environment of SMBHs. In a process that is not understood well, the interaction of the accretion disk and the SMBH drives the formation of jets. The blazar spectra therefore carry the information about the accretion and the circumnuclear environment, hence on the evolution of the environment of SMBHs over cosmic time. Indeed, a general characteristic of radio-loud quasars at high redshift is an apparent absorption in excess of the Galactic value \citep{fiore98, reeves00}. The nature of this X-ray spectral feature is still controversial. Obtaining precise spectra of blazars allows us to understand the physics at work in jets and derive information on the environment where the jet propagates. This in turn can confirm or deny evolutionary scenarios of blazars. 

PKS 0537-286, at z = 3.104 \citep{wright78}, is one of the most luminous known
high-redshift quasars. First detected at radio frequencies \citep{bolton75}, it was observed
at X-rays by the Einstein observatory \citep{zamorani81} and later studied with ASCA
\citep{cappi97,siebert96}, ROSAT \citep{fiore98}, XMM \citep{reeves01}
and {\it Swift} \citep{sambruna07}. These observations showed an extremely luminous quasar
(L$_{x}$=10$^{47}$ erg s$^{-1}$ in the 0.1 - 2 keV range) with a particularly hard
spectrum $(\Gamma = 1$ measured by {\it Swift}/BAT) and a weak iron K emission line and reflection features. Moreover, \citet{sowards-emmerd04} 
identified PKS 0537-286 as probable counterpart of the EGRET source 3EG
J0531-2940 \citep{hartman99}.

In this paper we report on a multiwavelength campaign on PKS 0537-286 ranging from the near-IR/optical to hard X-ray energies.  In S. 2 we describe the observations of the 2006 and 2008 multiwavelength campaigns; and in S. 3 the archival 2005 XMM observations; in S. 4  we report the results of our multiwavelenght observations and of the XMM archival data; in S. 4 we discuss the results. 

\section{Observations}

\subsection{The 2006 to 2008 multiwavelength campaign}

The source was monitored by INTEGRAL. Since the
{\it Swift}/BAT light curve over 9 months of exposure \citep{ajello08b} showed
a constant flux level, the INTEGRAL pointings were proposed in non-contiguous time intervals. ISGRI detected the source
at 5.6 $\sigma$.
In order to cover simultaneously the X-ray and UV ranges we requested
{\it Swift} ToOs. These observations were scheduled on October 27$^{th}$, 30$^{th}$, 31$^{st}$ 2006 and further
{\it Swift}/XRT observations were executed on February 10$^{th}$, 12$^{th}$ 2008. Simultaneously to the {\it Swift}
pointings REM, GROND, and Palomar 60-inch telescope performed photometry.
RXTE PCA observed PKS 0537-286 on November 1$^{st}$ 2006 in the 3 -
20 keV band. Details of the multiwavelength campaigns can be found in Tables ~\ref{tab:uvot} to ~\ref{tab:xray}.

\subsubsection{UV to optical observations}

UVOT. -- PKS 0537-286 was targeted in the U-band with the UV-Optical Telescope UVOT \citep{roming05} onboard the {\it Swift} satellite. Observations were performed in three epochs in 2006 October and two epochs in 2008 February (Table~\ref{tab:uvot}). Here, we use the standard pipeline reduced image products, co-added and exposure corrected within the XIMAGE\footnote{See http://heasarc.gsfc.nasa.gov/docs/xanadu/ximage/ximage.html} environment. 

% begin table UVOT
\begin{table}
\begin{center}
\caption{Log of UVOT observations. Flux density in units of $10^{-14}$ erg cm$^{-2}$ s$^{-1}$ $\AA^{-1}$\label{tab:uvot}}
\begin{tabular}{ccc}
\hline\hline
UTC start date & exposure (ks) & U-band flux density \\
\hline
2006-10-27 05:55:49 & 1.007  & $(4.6\pm1.0)$\\
2006-10-30 06:15:23 & 1.090  & $(5.9\pm1.0)$\\ 
2006-10-31 06:21:27 & 1.084  & $(4.2\pm1.0)$\\
2008-02-10 16:22:46 & 0.484  & $(5.6\pm1.2)$\\
2008-02-12 16:24:40 & 0.925  & $(3.9\pm1.1)$\\
\hline
\end{tabular}
\end{center}
\end{table}

Palomar. -- $B$ and $V$-band observations with the robotic Palomar 60-inch telescope \citep{cenko06} were  carried out in eight nights between 2006 Oct 26.4 UT and  2006 Nov 2.9 UT (Table ~\ref{tab:p60}). Data were reduced with standard IRAF\footnote{IRAF is distributed by the National Optical Astronomy Observatory, which is operated by the Association for Research in Astronomy, Inc. under cooperative agreement with the National Science Foundation.} routines \citep{tody93}. Photometric calibration was performed relative to the USNO-B1 catalog\footnote{See http://www.nofs.navy.mil/data/fchpix}, which leads to a systematic contribution to the photometric uncertainties of $\sim$ 0.5 mag.

%begin table P60
\begin{table}
\begin{center}
\caption{Log of Palomar observations. $^{a}$ statistical errors only.\label{tab:p60}}
\begin{tabular}{lccccc}
\hline\hline
MJD & UTC & band & Mag & Err$^{a}$ & Seeing \\
\hline
54033.40301 & 2006-10-25 09:40:20 &  B  &     19.58 &   0.02 &    2"4 \\
54033.41492 & 2006-10-25 09:57:29 &  V  &     18.90 &   0.01 &    2"0 \\
54034.42265 & 2006-10-26 10:08:36 &  B  &     19.31 &   0.03 &    3"8 \\
54034.43479 & 2006-10-26 10:26:05 &  V  &     18.86 &   0.02 &    2"5 \\
54035.41863 & 2006-10-27 10:02:49 &  B  &     19.59 &   0.03 &    3"7 \\
54035.43105 & 2006-10-27 10:20:42 &  V  &     18.99 &   0.02 &    3"1 \\ 
54036.41601 & 2006-10-28 09:59:03 &  B  &     19.68 &   0.02 &    2"8 \\
54036.42706 & 2006-10-28 10:14:57 &  V  &     18.86 &   0.02 &    2"5 \\
54038.40941 & 2006-10-30 09:49:33 &  B  &     19.55 &   0.03 &    3"7 \\
54038.42128 & 2006-10-30 10:06:38 &  V  &     18.96 &   0.01 &    3"4 \\
54039.40719 & 2006-10-31 09:46:21 &  B  &     19.64 &   0.02 &    2"7 \\
54039.41966 & 2006-10-31 10:04:18 &  V  &     18.89 &   0.01 &    2"3 \\
54040.40498 & 2006-11-01 09:43:10 &  B  &     19.57 &   0.02 &    2"0 \\
54040.41697 & 2006-11-01 10:00:26 &  V  &     18.67 &   0.06 &    1"7 \\
\hline
\end{tabular}
\end{center}
\end{table}

REM. -- $V, R, J, I, H$ observations were obtained with the Rapid Eye Mount (REM) \citep{chincarini03} in October and November 2006 (Table~\ref{tab:rem}). The REM data were reduced following standard procedures \citep{dolcini05}.

%begin table REM
\begin{table}
\begin{center}
\caption{Log of REM observations.\label{tab:rem}}
\begin{tabular}{lccccc}
\hline\hline
MJD& UTC & Filter & Mag & Err \\
\hline
54026.20225 & 2006-10-18 04:51:14    &  V  &     18.25 &   0.08\\
54036.14096 & 2006-10-28 03:22:58    &  V  &    	18.83 &   0.18\\
54038.24992 & 2006-10-30 05:59:53    &  V  &    	18.50 &   0.08\\
54039.22512 & 2006-10-31 05:24:10    &  V  &    	18.46 & 	0.10\\
54040.25737 & 2006-11-01 06:10:36    &  V  &    	18.39 & 	0.09\\
54026.24084 & 2006-10-18 05:46:48    &  R  &    	17.67 & 	0.08\\
54036.17956 & 2006-10-28 04:18:33    &  R  &    	17.93 & 	0.09\\
54038.17376 & 2006-10-30 04:10:12    &  R  &    	18.36 & 	0.12\\
54039.26192 & 2006-10-31 06:17:09    &  R  &    	18.55 & 	0.09\\
54040.29142 & 2006-11-01 06:59:38    &  R  &    	18.48 & 	0.10\\
54026.27942 & 2006-10-18 06:42:21    &  I  &    	17.15 & 	0.09\\
54038.21236 & 2006-10-30 05:05:47    &  I  &    	17.74 & 	0.12\\
54039.20966 & 2006-10-31 05:01:54    &  I  &    	17.77 & 	0.13\\
54040.23736 & 2006-11-01 05:41:47    &  I  &    	17.77 & 	0.11\\
54026.18431 & 2006-10-18 04:25:24    &  J  &    	17.05 & 	0.18\\
54036.12300 & 2006-10-28 02:57:07    &  J  &    	17.68 & 	0.20\\
54038.19724 & 2006-10-30 04:44:01    &  J  &    	17.66 & 	0.17\\
54039.17244 & 2006-10-31 04:08:18    &  J  &    	17.65 & 	0.14\\
54040.18487 & 2006-11-01 04:26:12    &  J  &    	17.73 & 	0.14\\
54026.22602 & 2006-10-18 05:25:28    &  H  &    	16.43 & 	0.16\\
54036.16471 & 2006-10-28 03:57:10    &  H  &    	16.77 & 	0.14\\
54038.15901 & 2006-10-30 03:48:58    &  H  &    	17.41 & 	0.11\\
54039.21766 & 2006-10-31 05:13:25    &  H  &    	17.10 & 	0.08\\
\hline
\end{tabular}
\end{center}
\end{table}

GROND. -- The Gamma-Ray Burst Optical and Near Infrared Detector (GROND) is a 7-channels imager primarily designed for fast follow-up observations of GRB afterglows \citep{greiner08}. The use of dichroic beam splitters allows simultaneous imaging in seven bands, $g^\prime r^\prime i^\prime z^\prime$ (similar to the Sloan system) and $J H K$ (400 nm - 2310 nm). GROND is mounted on the 2.2 m ESO/MPI telescope on LaSilla/Chile since April 2007. The observations were done in 2 epochs in February 2008 (Table~\ref{tab:grond}).

%begin table GROND
\begin{table}
\begin{center}
\caption{Log of GROND observations.\label{tab:grond}}
\begin{tabular}{lcccc}
\hline\hline
Date (UTC)& Filter & Mag & Err \\
\hline
2008-02-14 02:11:05     &  g$^{\prime}$  &    19.25 &   0.03\\
2008-02-14 02:11:05     &  r$^{\prime}$  &  18.89 &   0.04\\
2008-02-14 02:11:05     &  i$^{\prime}$  &    18.86 &   0.04\\
2008-02-14 02:11:05     &  z$^{\prime}$  &    18.88 & 	0.05\\
2008-02-14 02:11:05     &  J  &    	18.81 & 	0.06\\
2008-02-14 02:11:05     &  H  &    	18.76 & 	0.07\\
2008-02-14 02:11:05     &  K  &    	18.58 & 	0.06\\
2008-02-14 05:21:55     &  g$^{\prime}$ &    	19.25 & 	0.03\\
2008-02-14 05:21:55     &  r$^{\prime}$ &    	18.89 & 	0.04\\
2008-02-14 05:21:55     &  i$^{\prime}$ &    	18.85 & 	0.04\\
2008-02-14 05:21:55     &  z$^{\prime}$ &    	18.82 & 	0.05\\
2008-02-14 05:21:55     &  J  &    	18.85 & 	0.06\\
2008-02-14 05:21:55     &  H  &    	18.75 & 	0.07\\
2008-02-14 05:21:55     &  K  &    	18.66 & 	0.06\\
\hline
\end{tabular}
\end{center}
\end{table}

\subsection{X-ray observations}

INTEGRAL. -- The INTEGRAL satellite observed the source for 1 Ms with its imager IBIS/ISGRI \citep{lebrun03} which operates in the range of 17 - 1000 keV. The long exposure of 1 Ms was carried out in non-consecutive pointings. These observations were performed with a rectangular 5 $\times$ 5 dithering pattern and reported in Table~\ref{tab:ibis}. We used the standard Off-line Science Analysis (OSA - \citealt{courvoisier03}) software version 7.0 for the ISGRI analysis. Using the detected count rate we performed data cleaning. Most recent matrices available for standard software (isgri\_arf\_rsp\_0025.fits) were used for spectral analysis. We used 296 science windows for a total amount of 981 ksec of exposure time on the source. The source is detected with ISGRI in the 17 - 55 keV energy range with a significance of 5.6 sigma. Data screening was performed according to the median count rate with respect to each science window and its distribution. Because of the low detection level a spectrum could not be extracted.
The adopted dithering pattern resulted in a limited coverage of the source by JEM-X (Joint European Monitor for X rays - 
\citealt{lund03}). For 77\% of the time the source is outside the FOV. For the remaining time the source is at angles $<$ 4$^{\circ}$ from the axis. 
The total JEM-X useful exposure time is only 230 ks, insufficient for a detection.

%begin table ibis
\begin{table}
\begin{center}
\caption{Log of IBIS/ISGRI observations.\label{tab:ibis}}
\begin{tabular}{cccc}
\hline\hline
Rev. & Start date (UTC)& End date (UTC) & Exp (sec) \\
\hline
0493 & 2006-10-26 23:31:06 & 2006-10-29 04:19:28 & 180423\\
0494 & 2006-10-29 22:36:24 & 2006-11-01 12:51:01 & 202377\\
0549 & 2007-04-12 23:33:26 & 2007-04-15 03:02:26 & 170054\\
0550 & 2007-04-15 17:04:49 & 2007-04-18 01:32:37 & 188714\\
0552 & 2007-04-22 07:17:30 & 2007-02-24 01:23:05 & 136084\\
0558 & 2007-05-09 11:34:04 & 2007-05-10 21:03:40 & 115482\\ \hline
\end{tabular}
\end{center}
\end{table}

Swift/BAT. -- The Burst-Alert Telescope \citep{barthelmy05} on board the {\it Swift} satellite is a coded-mask telescope operating in the 15 - 200 keV energy range. Thanks to its large field of view, BAT surveys up to 80\% of the sky every day. We selected all observations of PKS 0537-286 comprised in the time span January 2006 - March 2008
. The data were processed using HEAsoft 6.3 and following the recipes presented in \citet{ajello08b}.The spectrum and the light curve of PKS 0537-286 were extracted using the method presented in \citet{ajello08a}.

RXTE. -- The Rossi X-ray Timing Explorer (RXTE) observed PKS 0537-286 for 9.9 ksec, starting on 2006 November 1. We reduced the data using the standard reduction script REX included in the HEAsoft6.0.4 package.  The analysis was restricted to standard 2 binned data of layer 1 of the Proportional Counting Unit 2 (PCU2) of the Proportional Counter Array (PCA) from 3 - 20 keV. We used PCARSP to produce the response matrix for the dataset.

Swift/XRT. --  The X-Ray Telescope (XRT) \citep{burrows05} on board {\it Swift} observed the blazar in October 2006 and February 2008. The monitoring campaign consisted of  five observations. Data processing, screening and filtering were done using the FTOOL xrtpipeline included in the HEAsoft 6.3 distribution.

%begin table xray
\begin{table}
\begin{center}
\caption{Log of X-ray observations.\label{tab:xray}}
\begin{tabular}{ccc}
\hline\hline
Instrument & Start date (UTC)& Exp (sec)\\
\hline
XRT & 2006-10-27 05:55:53 & 3000\\
XRT & 2006-10-30 06:15:21 & 3800\\
XRT & 2006-10-31 06:21:26 & 3400\\
XRT & 2008-02-10 16:22:48 & 6700\\
XRT & 2008-02-12 16:24:38 & 5200\\
RXTE & 2006-11-01 00:06:40 & 7200\\
XMM & 2005-03-20 16:05:56 & 80000\\
BAT & 2006 - 2008 & 2-years\\
\hline
\end{tabular}
\end{center}
\end{table}

\section{XMM-Newton archive data}

We searched for observations of PKS 0537-286 in the XMM-Newton archive 
and found a 80 ks observation of the source (OBS-ID 0206350201) still unpublished.
The source has been observed with the EPIC imagers.
Data were linearized by applying the most recent calibrations
with the software XMM-SAS version 9.0. We have extracted the light-curve of the 
background in the 10-15 keV energy range and excluded with a 
3$\sigma$ clipping technique a time interval of flaring particle background. 
Since the spectrum of many soft-proton flares may be extremely soft,
we repeated the light-curve screening in the 0.3-10 keV energy band and 
excluded time interval with high background flux. 
As a result, most of the observing time has been lost because of particle
flares and the final sum of the good time interval yields an effective exposure of $\sim$8 ks.

By using the same software package we extracted the EPIC-PN spectrum of the source
in a circle with a radius of 
40 arcsec. The background has been evaluated on a similar area of the same PN-CCD, but far
from source readout features. As a result the source spectrum contains $\sim$3700 net counts in the 
0.6-10 keV band which ensure one order of magnitude better statistics in comparison with the {\it Swift} data.

We further analyzed the XMM-Newton data of PKS 0537-286 observed during orbit 51 (but with latest response) 
in 1999 and reproduce the fit results as reported in \citet{reeves01}. 

\section{Results}

\subsection{Variability analysis}

Each single optical light curve obtained with REM, GROND and P60 does not show significant variations. Whenever, if at a given date, more than one REM, GROND or Palomar measurements are available, then the average flux is taken. Among the observations performed with the different optical instruments (this means
among long time scales) the source showed variability as can be seen in Tables ~\ref{tab:p60}, ~\ref{tab:rem}, ~\ref{tab:grond}. Within each individual {\it Swift}/XRT pointing just very marginal flux variation in the XRT spectrum was found (Table~\ref{tab:fit}). Considering the large uncertainties among the five pointings, the flux amplitude is $\sim$8$\%$. The BAT light curve is consistent with being constant over the 2 two years of the survey.

\subsection{{\it Swift} and RXTE spectra}

The fit results of the single spectra of each single instrument are shown in Table~\ref{tab:fit}. The spectra were fitted using XSPEC 12 and the latest available response matrices for calibration. The spectra are exceptionally hard.
{\it Swift}/BAT detected, in the 15 - 150 keV range, PKS 0537-286 with a flux of $\sim$ 3 $\times$ 10$^{-11}$ erg cm$^{-2}$ s$^{-1}$ during the two years of the survey. The low Galactic N$_{H}$-value (2.0 $\times$ 10$^{20}$ cm$^{-2}$ \citet{kalberla05}) does not affect the BAT energy range, and we do not see evidence of further absorption or deviations from a single power law.

Since the lower limit of the energy range is around 2 keV also the RXTE spectrum is not affected by absorption features. Therefore also this spectrum is well represented by a simple power law.

The best fit results ($\chi^{2}_{red}$ $<$ 0.9) for each single {\it Swift}/XRT spectrum is given by an absorbed power-law model, with absorption parameter free to vary. Given the large fit uncertainties, reflecting the poor quality of these spectra, it is impossible to fully constrain the obtained parameters.

Since we would like to assess the spectral characteristics with the best possible S/N, we have joined the average BAT, XRT and RXTE spectra and fitted the combined spectrum. The fit result are listed in Table ~\ref{tab:fit}.
 
In the case of the co-added XRT, RXTE and BAT spectra we model the data with a power law with Galactic absorption controlled by the hydrogen column density. The large errors of the fitted  parameters do not allow to constrain the absorption parameter.

%begin table fit
\begin{table}
\begin{minipage}[][4.8cm][c]{\columnwidth}
%\begin{minipage}[t]{\columnwidth}
\begin{center}
\caption{X-ray fit results.\label{tab:fit}}
\renewcommand{\footnoterule}{}
\begin{tabular}{ccccccc}
\hline\hline
Inst. & obs epoch & fit model & N$_{H}$ & $\Gamma$ & norm & flux$^{a}$\\
(1) & (2) & (3) & (4) & (5) & (6) & (7)\\
\hline
XRT & 2006-10-27 & abs pl & 0.08$^{0.26}_{0.00}$ & 1.3$^{1.7}_{0.9}$ & 3.9$^{6.2}_{2.6}$ & 2.4$^{2.8}_{2.0}$\\
XRT & 2006-10-30 & abs pl & 0.15$^{0.20}_{0.06}$ & 1.7$^{2.2}_{1.4}$ & 6.9$^{9.6}_{5.0}$ & 2.6$^{3.0}_{2.0}$\\
XRT & 2006-10-31 & abs pl & 0.10$^{0.22}_{0.01}$ & 1.5$^{1.8}_{1.2}$ & 5.1$^{7.3}_{3.6}$ & 2.4$^{2.8}_{1.9}$\\
XRT & 2008-02-10 & abs pl & 0.05$^{0.12}_{0.00}$ & 1.3$^{1.5}_{1.1}$ & 4.2$^{5.4}_{3.3}$ & 2.6$^{2.8}_{2.1}$\\
XRT & 2008-02-12 & abs pl & 0.06$^{0.18}_{0.00}$ & 1.3$^{1.5}_{1.0}$ & 3.6$^{5.1}_{2.7}$ & 2.3$^{2.7}_{2.0}$\\
RXTE & 2006-11-01 & pl & - & 1.3$^{1.7}_{1.0}$ & 5.7$^{1.3}_{0.0}$ & 6.3$^{7.2}_{4.6}$\\
BAT & 2006 - 2008 & pl & - & 1.5$^{1.9}_{1.0}$ & 14$^{71}_{0.0}$ & 0.27$^{0.34}_{0.21}$\\
joint & - & abs pl & 0.05$^{0.08}_{0.02}$ & 1.3$^{1.4}_{1.2}$ & 4.2$^{4.6}_{3.8}$ & - \\
\hline
\end{tabular}
\end{center}
\end{minipage}
Explanation of columns: (1)=instruments; (2)=observation date; (3)=fit model; (4)=column density (units of 10$^{22}$ cm$^{-2}$), (5)=spectral index; (6)=normalization (units of 10$^{-4}$ ph keV$^{-1}$ cm$^{-2}$); (7)=flux units are 10$^{-12}$ erg cm$^{-2}$ s$^{-1}$ for all instruments

$^{a}$fluxes are computed in the following energy ranges for the different instruments: XRT 0.6 - 6.0 keV , RXTE  3 - 20 keV, BAT 15 -150 keV, joint 0.6 - 150 keV.

\end{table}
 
\subsection{XMM spectra}

%begin table fit
\begin{table*}
\begin{minipage}[][3.5cm][c]{\textwidth}
\begin{center}
\caption{Fit results of X-ray XMM spectra.\label{tab:xmm}}
\renewcommand{\footnoterule}{}
\begin{tabular}{ccccccccc}
\hline\hline
inst & model & N$_{H}$ & $\Gamma$ & norm & KT & norm BB & N$_{H}^{z}$ & $\chi^{2}(dof)$\\
(1) & (2) & (3) & (4) & (5) & (6) & (7) & (8) & (9)\\
\hline
PN-M1-M2 & ab*pl & 2.0 & 1.15$^{1.16}_{1.14}$ & 4.5$^{4.6}_{4.4}$ & -  & - & - & 1110(1013)\\
PN-M1-M2 & ab*pl & 4.5$^{5.1}_{3.9}$ & 1.24$^{1.26}_{1.21}$ & 5.0$^{5.1}_{4.9}$ & -  & - & - & 1050(1012)\\
PN-M1-M2 & ab(ab$_{z}$*pl) & 2.0 & 1.22$^{1.24}_{1.20}$ & 4.8$^{4.9}_{4.7}$ & -  & - & 4.8$^{5.9}_{3.8}$ & 1047(1012)\\
PN-M1-M2 & ab(pl+bb) & 2.0 & 1.08$^{1.09}_{1.06}$ & 3.9$^{4.0}_{3.7}$ & 0.35$^{0.40}_{0.32}$  & 2.4$^{3.1}_{1.9}$ & - & 1057(1011)\\
\hline
\end{tabular}
\end{center}
\end{minipage}
Explanation of columns: (1)=instruments; (2)=fit model; (3)=column density (units of 10$^{20}$ cm$^{-2}$), (4)=spectral index; (5)=normalization (units of 10$^{-4}$ ph keV$^{-1}$ cm$^{-2}$); (6)=photon temperature; (7)=blackbody normalization; (8)=absorption parameter at the source (units of 10$^{21}$ cm$^{-2}$); (9)=statistical fit result.
\end{table*}

First we have checked the consistency of the single X-ray spectra among the 3 EPIC detectors within the single observations by fitting power-law model as in \citet{reeves01}. Our re-analysis of the observation in 1999 (orbit 51) confirms the results reported by \citet{reeves01}, where a simple power-law model fits the spectrum to a spectral index $\Gamma$ = 1.22 $\pm$ 0.08 with $\chi^{2}$(dof) = 983(808). By fitting these data to an absorbed power law, the absorption parameter is consistent with the Galactic hydrogen column density. Therefore no absorption in addition the the Galactic $N_{H}$ is required, confirming once more the results in \citet{reeves01}.

\begin{figure}
\centering
\includegraphics[height=6.5cm,keepaspectratio=true]{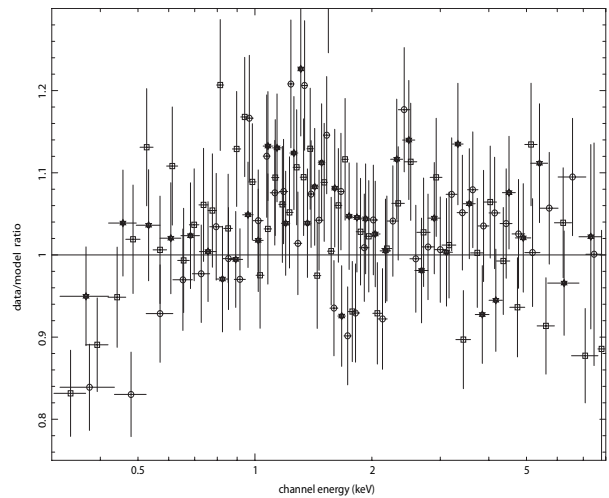}
\caption{Data-to-model ratio for PKS 0537-286 as observed by XMM in 2005. The fitted model is a power law (including Galactic absorption) in the 0.3 - 8.0 keV energy range. The model is inadequate. The data are rebinned for demonstration to have at least a significant detection of 15 $\sigma$. The pn data are represented by squares, while circles and stars represent MOS-1 and MOS-2, respectively.\label{fig:fig01}}
\end{figure}

\begin{figure}
\centering
\includegraphics[height=6.5cm,keepaspectratio=true]{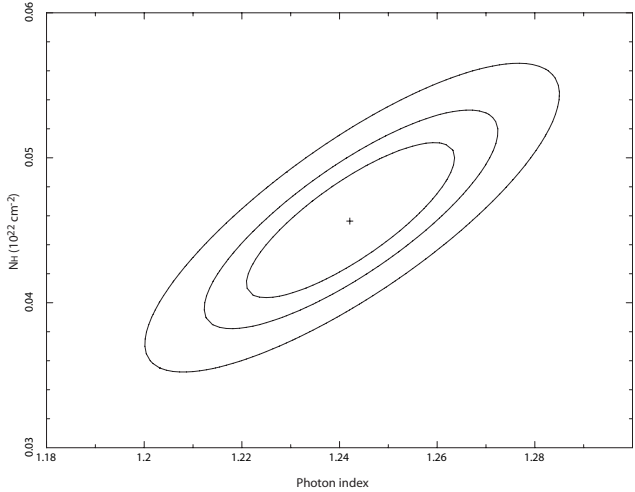}
\caption{Confidence level contour plot of the spectral index and the absorption parameter at 68$\%$, 90$\%$ and 99$\%$ confidence levels of the spectrum extracted from the XMM observation of 2005. \label{fig:fig02}}
\end{figure}

Contrary to the observation in 1999, the spectra derived from the observation in 2005 show an apparent absorption in excess to the Galactic value at $\sim$7 $\sigma$. Figure ~\ref{fig:fig01} shows the data-to-model ratio of the best-fit power law modified by the Galactic absorption (with $\chi^{2}$(dof) = 1110(1013)). Besides falling to $\sim$ 0.8 for the softest energies, the residuals exhibit emission in excess between 1 - 2 keV. This is similar to what is found by \citet{yuan05}, \citet{worsley04a} and \citet{tavecchio07} for other high-reshift blazars. Therefore is the spectrum modeled by an absorbed power law with $\chi^{2}$(dof) = 1015(1012). Figure ~\ref{fig:fig02} shows the confidence level contour plot of the absorption in excess to the Galactic value N$_{H}$ and the spectral index. In order to reproduce this clear spectral feature, we add another component to the base line model (wabs*powerlaw) and fit the photoelectric absorption component fixed to the Galactic value convolved with the intrinsic absorption component and a power law (wabs(zwabs*powerlaw)). The resulting intrinsic absorption value is N$^{z}_{H}$ $\sim$ 4.8$\times$10$^{21}$ cm$^{-2}$. Tabel ~\ref{tab:xmm} shows the fit results of the co-added PN, MOS-1 and MOS-2 spectrum.

\begin{figure}
\centering
\includegraphics[height=7cm,keepaspectratio=true]{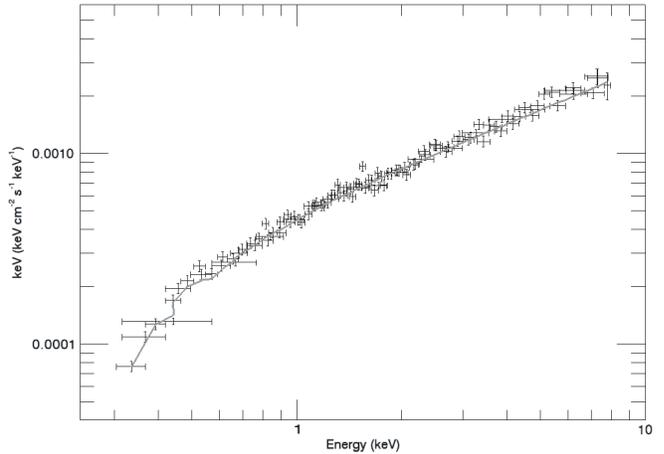}
\caption{Unfolded XMM spectrum between 0.3 and 8 keV in the $\nu$$F_{\nu}$ space. The spectrum is modeled with an power law and absorption parameter free to vary.\label{fig:fig03}}
\end{figure}

A fit of the spectrum with a power law and black body, modified by Galactic absorption as suggested in \citet{celotti07}, gives an acceptable fit ($\chi^{2}_{red} = 1.04$). But there is only marginal improvement of the $\chi^{2}_{red}$ (1.09) obtained fitting  the power law with absorption fixed to the Galactic value. We further show that the same spectrum is modeled with $\chi^{2}_{red} = 1.04$ by a power law and free absorption parameter. Figure ~\ref{fig:fig03} shows the unfolded spectrum fitted to the latter model in E$^{2}f(E)$ space.

A systematic spectral flattening of high-redshift blazars at low X-ray energies was found previously \citep{fiore98,reeves00}. Similarly \citet{worsley04b} detected a convex spectrum for the blazar PMN J0525-3343 at z = 4.4, and argued that it is most likely produced by a warm (ionized) absorber intrinsic to the source. We probed this hypothesis for PKS 0537-286 by fitting a power law and the contribution of ionized media intervening between source and the observer that modify the source flux in an energy-dependent way\footnote{The model is a convolution of three XSPEC models, namely powerlaw, absori and wabs fixed to the Galactic value.}. The warm absorber hypothesis results in an absorbing ionization state $\xi$ = 150$^{450}_{0.0}$ erg cm$^{-2}$  s$^{-1}$.

The XRT flux in the 0.6 - 6.0 keV band is systematically lower than the flux in the same energy band seen in XMM. Moreover, the spectral index of the XMM spectrum is slightly harder than the one of the XRT spectra. This suggests significant X-ray flux and spectral variability between 2005 and 2006-2008, so that we have not attempted any joint fit of the RXTE, {\it Swift} and XMM spectra.

\section{The spectral energy distribution}

%begin table SED
\begin{table*}[!t]
\begin{minipage}[][3.5cm][c]{\textwidth}
\centering
\renewcommand{\footnoterule}{}
\caption{List of parameters used to construct the theoretical SED.}
\begin{tabular}{llllllllllll}
\hline
\hline
$R_{\rm diss}$ &$M$ &$R_{\rm BLR}$ &$P^\prime_{\rm i}$ &$L_{\rm d}$ &$B$ &$\Gamma$ &$\theta_{\rm v}$
    &$\gamma_{\rm b}$ &$\gamma_{\rm max}$ &$s_1$  &$s_2$\\
~(1)&(2) &(3) &(4) &(5) &(6) &(7) &(8) &(9) &(10) &(11) &(12)\\
\hline   
4.2 (2)   &7 &1.3$\times$10$^{3}$ &0.3   &168  (0.16)  &0.1    &18   &3   &40  &10$^{4}$   &--1   &2.6\\ 
3.2 (1.5) &7 &916   &0.2  &84 (0.08) &0.09  &18   &3   &25  &10$^{4}$   &--1   &2.6\\ 
\hline
\hline 
\end{tabular}
\label{tab:parameters}
\end{minipage}
Explanation of columns:
(1) = dissipation distance in units of $10^{18}$ cm and (in parenthesis) in units of $10^{3}$ $R_{\rm S}$;
the (spherical) emitting size is ten times smaller;
(2) = black hole mass in 10$^{9}$ solar masses;
(3) = size of the BLR in units of $10^{15}$ cm;
(4) = power injected in the blob calculated in the comoving frame, in units of $10^{45}$ erg s$^{-1}$; 
(5) = accretion-disk luminosity in units of $10^{45}$ erg s$^{-1}$ and (in parenthesis) in units of $L_{\rm Edd}$;
(6) = magnetic field in Gauss;
(7) = bulk Lorentz factor at $R_{\rm diss}$;
(8) = viewing angle in degrees;
(9) and (10) = break and maximum random Lorentz factors of the injected electrons;
(11) and (12) = slopes of the injected electron distribution ($Q(\gamma)$) below and above $\gamma_{\rm b}$.
\end{table*}

\begin{figure*}
\centering
\includegraphics[height=17cm,keepaspectratio=true]{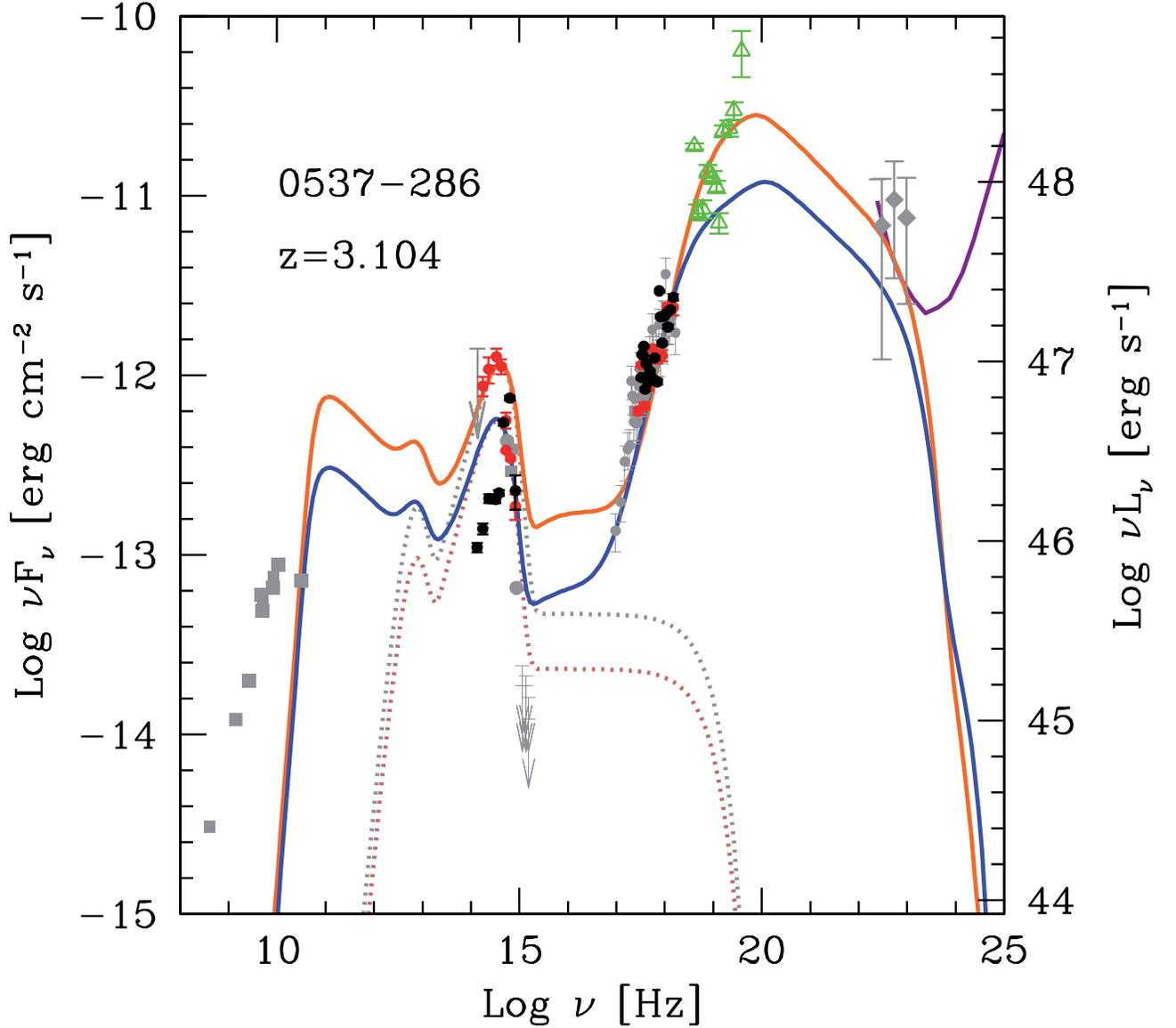}
\caption{Observed spectral energy distribution of PKS 0537-286 on 27 October 2006 (black filled circles) and 10 February 2008 (red filled circles). The filled green triangles represent the {\it Swift}/BAT spectrum extracted from the 2 years survey. The data are corrected for Galactic extinction. Gray points are archival data. The dotted line shows the single components to the SED from the IR torus, the disk and the X-ray corona. The purple solid line represents the 5 $\sigma$ sensitivity of Fermi after one year mission time. The jet model \citet{ghisellini09} for the energy distribution for 27 October 2006 (orange solid curve) and 10 February 2008 (blue solid curve) are overplotted.\label{fig:fig04}}
\end{figure*}

We constructed two multiwavelength spectra using the data of the first and fourth epochs (corresponding to the XRT pointings on October 27$^{th}$ 2006 and on February 10$^{th}$ 2008) of our campaign, and adding to these non-simultaneous multiwavelength data, including  an EGRET detection in  three energy bins in the GeV range.   The optical and UV data were corrected for Galactic absorption using E(B-V) = 0.025 mag \citep{schlegel98}, and converted into fluxes. 
The optical flux at  wavelengths shorter than $\sim$5000 \AA\ is significantly suppressed by the Ly $\alpha$ intervening absorption \citep{weymann81}. Therefore is the sharp increase from UV towards optical band not surprising, as the Ly $\alpha$ forest is shifted along lower frequencies due to the high-redshift. PKS 0537-286 belongs to the FSRQ subclass of blazars. The typical two hump structure is evident in the two SEDs. The IC-peak can be constrained by the BAT data points and EGRET data to be located at a few GeV. The synchrotron peak is somewhat difficult to determine.

We model the data according to \citet{ghisellini09}, assuming a one-zone, homogeneous synchrotron and
inverse-Compton model. 
The ejecta move within the jet seen under an angle and bulk Lorentz 
factor $\Gamma$. 
The homogeneous magnetic field is tangled throughout the emitting region.
This is located at a distance $R_{\rm diss}$ from the black hole,
has a cross sectional radius $R=0.1 R_{\rm diss}$ and a 
(comoving) width $\Delta R^\prime=R$.
The particle distribution is derived through a continuity equation,
and assume that particles are injected for 
a finite time, of duration $t_{inj}=\Delta R'/c$.
Radiative cooling and electron-positron pair production and emission
(not relevant in this case) are accounted for.
Although the considered continuity equation has no steady state solution,
we evaluate it at the time $t_{inj}$, when it reaches the maximum value,
corresponding to the maximum emitted luminosity.
The broad line region (BLR) clouds reprocess and re-isotropize 10\% 
of the disk luminosity.
Since we assume that the radius of the BLR is $\propto L_{\rm d}^{1/2}$,
the radiation energy density of the broad lines (within the BLR) as seen in the
comoving frame depends only on $\Gamma^2$: 
$U^\prime_{\rm BLR}\propto \Gamma^2 L_{\rm BLR}/R^2_{\rm BLR}= $const$\times \Gamma ^2$.
This component, as seen in the comoving frame, is approximated with a black body 
spectrum (see Tavecchio \& Ghisellini 2008).
We assume that the injected particle distribution is a smoothly broken power law,
with slopes $s_1$ and $s_2$ below and above 
the electron Lorentz factor $\gamma_{\rm b}$.
In the case of our source, radiative cooling is efficient, implying
that electrons of all energies have time to cool in $t_{inj}$.
As a consequence, the particular value of $s_1$ is ininfluent, since,
below $\gamma_{\rm b}$, the emitting particle distribution is
$\propto \gamma^{-2}$, while it becomes $\propto \gamma^{-(s_2+1)}$
above $\gamma_{\rm b}$.
We modeled the first and the fourth observing epochs which show 
the largest variation from near-IR to UV band. 
Using the detection in three energy
bins by EGRET, the IC peak was assumed to be between these
data points and the Swift/BAT spectrum. The results are shown in Figure ~\ref{fig:fig04} (solid orange and blue lines). The assumed parameters are reported in Table~\ref{tab:parameters}.
 
Table~\ref{tab:parameters-power} lists the power carried by the jet in the form of
radiation ($P_{\rm r}$), magnetic field ($P_{\rm B}$), electrons ($P_{\rm e}$) 
and protons ($P_{\rm p}$, assuming one proton per emitting electron). 
The powers are
calculated as 
\begin{equation}
P_i\, =\, \pi R^2 \Gamma^2 \beta c U_i
\end{equation}
where $U_i$ is the energy density of the $i^{th}$ component.

The optical black data points are not well represented by the model. Indeed, assuming the highest data point to be the peak of the accretion-disk component, then the model requires a factor $\sim$3 more in temperature. Being the relation between accretion-disk luminosity (L$_d$), accretion rate ($\dot{M}$) and temperature (T):
\begin{equation}
L_{d} \propto \dot{M} \propto T^{4},
\end{equation}
that, in turn would require a very high variation of the accretion rate (factor $\sim$80), at odds with the model. Therefore we decided to assume the peak of the accretion-disk component around the GROND $z^{\prime}$ data point, where a turn over in the data sequence can be seen and that can well reproduce the IC component.
The jet power and the observed non--thermal luminosity increase
as the disk luminosity increases, despite the fact that the
amount of inverse Compton cooling is constant (i.e. $U^\prime_{\rm BLR}$ is
constant).
The variations of the non--thermal bolometric luminosity are 
due to the different power injected in the form of relativistic electrons,
and the different Compton to synchrotron luminosity ratio is due
to the different assumed magnetic field.

%begin table SED powers
\begin{table}[!h]
\begin{minipage}[][3cm][c]{\columnwidth}
%\begin{minipage}[t]{\columnwidth}
\centering
\caption{List of parameters used to construct the theoretical SED.}
\renewcommand{\footnoterule}{}
\begin{tabular}{llll}
\hline
\hline
$\log P_{\rm r}$ &$\log P_{\rm B}$ &$\log P_{\rm e}$ &$\log P_{\rm p}$ \\ 
(1) &(2) &(3) &(4) \\
\hline   
46.5 &45.3 &46.6 &48.0  \\ 
46.2 &45.0 &46.4 &48.0 \\ 
\hline
\hline 
\end{tabular}
\label{tab:parameters-power}
\end{minipage}
Explanation of columns: (1), (2), (3) and (4): jet power in the form of radiation, magnetic field,
bulk motion of electrons and protons (assuming one proton
per emitting electron). These powers are derived quantities, not input parameters.
For all cases the X--ray corona luminosity $L_X=0.3 L_{\rm d}$.
Its spectral shape is assumed to be $\propto \nu^{-1} \exp(-h\nu/150~{\rm keV})$.
\end{table}

\section{Discussion}

It is unlikely that the apparent excess column density for PKS 0537-286 is of Galactic origin. It would require absorption by cold material with a column of  N$_{H}$ $\sim$ 2.5$\times$10$^{20}$ cm$^{-2}$ in addition to the Galactic level toward the direction of the source. The Galactic hydrogen column density derived within different energy bands (IR, radio) by \citet{schlegel98}, \citet{elvis89} and \citet{kalberla05} enforces the accuracy of the measured Galactic N$_{H}$. Also the intergalactic origin is highly improbable because the line of sight would have to intercept several systems with very high column density. The chance probability to intercept several very high column density Ly $\alpha$ absorption systems is very low (p $\sim$ 2$\%$) \citep{oflaherty97}. Furthermore, we exclude that the damped Ly $\alpha$ absorption systems (DLA) contribute significantly to the spectral absorption feature. In fact, the DLA at z = 2.974, associated with an intervening galaxy, has HI column density of $log N_{HI} = 20.3 \pm 0.1$ \citep{akerman05} and metallicity below half solar. This would give an absorption at soft X-ray energies caused by a column density of N$_{H}$ $<$ 10$^{20}$ cm$^{-2}$. The intrinsic absorber hypothesis is much debated. For this absorber we derived a high column density value of N$^{z}_{H}$ $\sim$ 4.8$\times$10$^{21}$ cm$^{-2}$. Under the warm absorber hypotesis we cannot constrain the ionization state of the possible absorbing material better than $\xi$ = 150$^{450}_{0.0}$ erg cm$^{-2}$ s$^{-1}$. Furthermore, the best evidence for warm absorbers, like the O absorption edge and the Fe M-shell, are down shifted outside our observed range (0.3 - 10 keV) being in the rest frame's energy band within 0.5 - 0.9 keV. The possibility for the warm absorber to give rise to absorption variability \citep{reeves00} is not confirmed in our case.

\begin{figure}
\centering
\includegraphics[height=6.5cm,keepaspectratio=true]{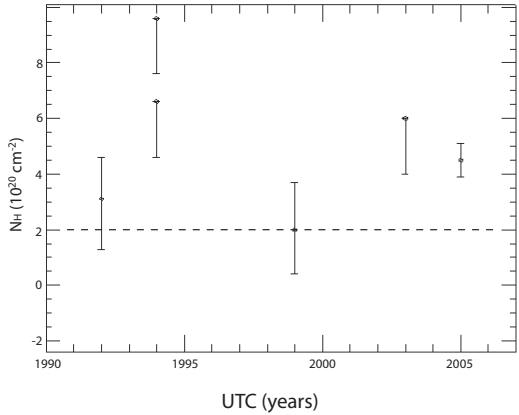}
\caption{Time variability of $N_{H}$ measured by several instruments. From left to right the instruments are: ROSAT \citep{cappi97}, two ASCA upper limits \citep{siebert96}, XMM (this work), Chandra 3 $\sigma$ upper limit(this work), XMM (this work). The dashed horizontal line delimits the Galactic hydrogen column density \citep{kalberla05}.\label{fig:fig05}}
\end{figure}

The excess absorption, also referred to as spectral flattening, has been reported several times for this source \citep{cappi97,fiore98,sambruna07}. Spectral flattening is in agreement with a general behavior seen in a number of high-redshift radio-loud quasars studies \citep{fiore98,reeves00}, where the required column density is up to $10^{22} cm^{-2}$, whereas low-redshift blazars show unabsorbed spectra. If the flattening is not a mere selection effect due to the high-redshift, then this could indicate an evolutionary scenario. In fact, the positive evolution of FSRQs \citep{dunlop90} foresees a bulk of FSRQs at high-redshift. PKS 0537-286 shows time variability of the apparent excess column density. Figure ~\ref{fig:fig05} shows the time variability of the column density over 7 years as measured by several instruments. Similarly \citet{schartel97} find variable absorption of X-rays for a high-redshift blazar on time scales of $< 1.5$ years. At present the origin of this absorption is unknown.

Inspection of the SEDs allows us to constrain the inverse Compton peak rather well, thanks to 
the exceptionally hard X-ray spectrum and the EGRET data.  
Moreover, our source is not included in the {\it Fermi} bright source list \citep{abdo09}, indicating that the 
upper limit is consistent. 
The model does not account very well for the shorter wavelength optical data, because of the Ly $\alpha$ drop.  An optical flux excess in the 2$^{nd}$ epoch is partially due to strong emission lines contaminating the GROND filters. 
The chosen parameters for the 1$^{st}$ observation epoch deviate from those obtained for the same source by \citet{sambruna07} and modeled using \citet{ghisellini02}. This is due to our accurately sampled data, that allow us to better constrain the theoretical SED.

\section{Conclusions}
We have conducted a multifrequency campaign on PKS 0537-286 on six different epochs. The source was monitored nearly simultaneously from optical to gamma-rays. In the optical - near-IR band the most evident flux variations were observed between the observations on October 27$^{th}$ 2006 and on February 10$^{th}$ 2008. The compiled SED for the latter two observation epochs allowed to derive the physical parameters of the source.
 
Also a new and detailed X-ray spectroscopic study of the XMM-Newton observation in 2005 of the source has been presented. The spectral flattening for this source at soft X-ray energies, reported previously by \citet{cappi97}, \citet{fiore98} and \citet{sambruna07}, is confirmed by the high signal-to-noise XMM spectra. This flux decrease at soft X-ray energies cannot be attributed to excess absorption in the Milky Way or to intervening material in the intergalactic medium along the line of sight to the source. Data suggest that the absorber can be intrinsic to the source. By using our data and searching in literature we have constructed a time curve of the $N_{H}$ for the source. We confirm the variability of the apparent excess absorption on time scale of years. A periodicity for this spectral feature cannot be inferred. Our result highlights once more the evidence for the existence of a population of superluminous blazars at high-redshift that show a soft X-ray cut-off. If the bulk at high-redshift of these type of sources is not only due to selection effect, then their existence is important in the context of cosmological evolution of SMBHs \citep{boettcher02}.

\begin{acknowledgements}

We thank Marcella Brusa for providing information on XMM-observation and Francesco
Longo for the help with data analysis.
E. Pian acknowledges financial support from ASI-INAF I/023/05/0
and ASI I/088/06/0A. TK acknowledges support by the DFG cluster of excellence 'Origin and
Structure of the Universe'. The anonymous referee is acknowledged for his helpful comments,
which improved the manuscript.
We also thank the INTEGRAL and the {\it Swift} team for the observations and the support.
AR acknowledges support through NASA grant NNX07AG33G. 

\end{acknowledgements}

\bibliographystyle{aa}
%\bibliography{biblio}

\begin{thebibliography}{54}
%\expandafter\ifx\csname natexlab\endcsname\relax\def\natexlab#1{#1}\fi

\bibitem[{{Abdo} {et~al.}(2009){Abdo}, {Ackermann}, {Ajello}, {Atwood},
  {Axelsson}, {Baldini}, {Ballet}, {Barbiellini}, {Bastieri}, {Baughman},
  {Bechtol}, {Bellazzini}, {Blandford}, {Bloom}, {Bonamente}, {Borgland},
  {Bouvier}, {Bregeon}, {Brez}, {Brigida}, {Bruel}, {Burnett}, {Caliandro},
  {Cameron}, {Caraveo}, {Casandjian}, {Cavazzuti}, {Cecchi}, {Charles},
  {Chekhtman}, {Chen}, {Cheung}, {Chiang}, {Ciprini}, {Claus}, {Cohen-Tanugi},
  {Colafrancesco}, {Collmar}, {Cominsky}, {Conrad}, {Costamante}, {Cutini},
  {Dermer}, {de Angelis}, {de Palma}, {Digel}, {do Couto e Silva}, {Drell},
  {Dubois}, {Dumora}, {Farnier}, {Favuzzi}, {Fegan}, {Ferrara}, {Finke},
  {Focke}, {Foschini}, {Frailis}, {Fuhrmann}, {Fukazawa}, {Funk}, {Fusco},
  {Gargano}, {Gasparrini}, {Gehrels}, {Germani}, {Giebels}, {Giglietto},
  {Giommi}, {Giordano}, {Giroletti}, {Glanzman}, {Godfrey}, {Grenier},
  {Grondin}, {Grove}, {Guillemot}, {Guiriec}, {Hanabata}, {Harding}, {Hartman},
  {Hayashida}, {Hays}, {Healey}, {Horan}, {Hughes}, {J{\'o}hannesson},
  {Johnson}, {Johnson}, {Johnson}, {Johnson}, {Kadler}, {Kamae}, {Katagiri},
  {Kataoka}, {Kerr}, {Kn{\"o}dlseder}, {Kocian}, {Kuehn}, {Kuss}, {Lande},
  {Latronico}, {Lemoine-Goumard}, {Longo}, {Loparco}, {Lott}, {Lovellette},
  {Lubrano}, {Madejski}, {Makeev}, {Massaro}, {Mazziotta}, {McConville},
  {McEnery}, {McGlynn}, {Meurer}, {Michelson}, {Mitthumsiri}, {Mizuno},
  {Moiseev}, {Monte}, {Monzani}, {Moretti}, {Morselli}, {Moskalenko}, {Murgia},
  {Nolan}, {Norris}, {Nuss}, {Ohsugi}, {Omodei}, {Orlando}, {Ormes}, {Ozaki},
  {Paneque}, {Panetta}, {Parent}, {Pelassa}, {Pepe}, {Pesce-Rollins}, {Piron},
  {Porter}, {Rain{\`o}}, {Rando}, {Razzano}, {Razzaque}, {Reimer}, {Reimer},
  {Reposeur}, {Reyes}, {Ritz}, {Rochester}, {Rodriguez}, {Romani}, {Ryde},
  {Sadrozinski}, {Sanchez}, {Sander}, {Saz Parkinson}, {Scargle}, {Schalk},
  {Sellerholm}, {Sgr{\`o}}, {Shaw}, {Smith}, {Smith}, {Spandre}, {Spinelli},
  {Starck}, {Strickman}, {Suson}, {Tajima}, {Takahashi}, {Takahashi}, {Tanaka},
  {Taylor}, {Thayer}, {Thayer}, {Thompson}, {Tibaldo}, {Torres}, {Tosti},
  {Tramacere}, {Uchiyama}, {Usher}, {Vilchez}, {Villata}, {Vitale}, {Waite},
  {Winer}, {Wood}, {Ylinen}, \& {Ziegler}}]{abdo09}
{Abdo}, A.~A., {Ackermann}, M., {Ajello}, M., {et~al.} 2009, \apj, 700, 597

\bibitem[{{Aharonian}(2000)}]{aharonian00}
{Aharonian}, F.~A. 2000, New Astronomy, 5, 377

\bibitem[{{Ajello} {et~al.}(2008{\natexlab{a}}){Ajello}, {Greiner}, {Kanbach},
  {Rau}, {Strong}, \& {Kennea}}]{ajello08b}
{Ajello}, M., {Greiner}, J., {Kanbach}, G., {et~al.} 2008{\natexlab{a}}, \apj,
  678, 102

\bibitem[{{Ajello} {et~al.}(2008{\natexlab{b}}){Ajello}, {Rau}, {Greiner},
  {Kanbach}, {Salvato}, {Strong}, {Barthelmy}, {Gehrels}, {Markwardt}, \&
  {Tueller}}]{ajello08a}
{Ajello}, M., {Rau}, A., {Greiner}, J., {et~al.} 2008{\natexlab{b}}, \apj, 673,
  96

\bibitem[{{Akerman} {et~al.}(2005){Akerman}, {Ellison}, {Pettini}, \&
  {Steidel}}]{akerman05}
{Akerman}, C.~J., {Ellison}, S.~L., {Pettini}, M., \& {Steidel}, C.~C. 2005,
  \aap, 440, 499

\bibitem[{{Antonucci}(1993)}]{antonucci93}
{Antonucci}, R. 1993, \araa, 31, 473

\bibitem[{{Barthelmy} {et~al.}(2005){Barthelmy}, {Barbier}, {Cummings},
  {Fenimore}, {Gehrels}, {Hullinger}, {Krimm}, {Markwardt}, {Palmer},
  {Parsons}, {Sato}, {Suzuki}, {Takahashi}, {Tashiro}, \&
  {Tueller}}]{barthelmy05}
{Barthelmy}, S.~D., {Barbier}, L.~M., {Cummings}, J.~R., {et~al.} 2005, Space
  Science Reviews, 120, 143

\bibitem[{{Bolton} {et~al.}(1975){Bolton}, {Shimmins}, \& {Wall}}]{bolton75}
{Bolton}, J.~G., {Shimmins}, A.~J., \& {Wall}, J.~V. 1975, Australian Journal
  of Physics Astrophysical Supplement, 34, 1

\bibitem[{{B{\"o}ttcher} \& {Dermer}(2002)}]{boettcher02}
{B{\"o}ttcher}, M. \& {Dermer}, C.~D. 2002, \apj, 564, 86

\bibitem[{{Burrows} {et~al.}(2005){Burrows}, {Hill}, {Nousek}, {Kennea},
  {Wells}, {Osborne}, {Abbey}, {Beardmore}, {Mukerjee}, {Short}, {Chincarini},
  {Campana}, {Citterio}, {Moretti}, {Pagani}, {Tagliaferri}, {Giommi},
  {Capalbi}, {Tamburelli}, {Angelini}, {Cusumano}, {Br{\"a}uninger}, {Burkert},
  \& {Hartner}}]{burrows05}
{Burrows}, D.~N., {Hill}, J.~E., {Nousek}, J.~A., {et~al.} 2005, Space Science
  Reviews, 120, 165

\bibitem[{{Cappi} {et~al.}(1997){Cappi}, {Matsuoka}, {Comastri}, {Brinkmann},
  {Elvis}, {Palumbo}, \& {Vignali}}]{cappi97}
{Cappi}, M., {Matsuoka}, M., {Comastri}, A., {et~al.} 1997, \apj, 478, 492

\bibitem[{{Celotti} {et~al.}(2007){Celotti}, {Ghisellini}, \&
  {Fabian}}]{celotti07}
{Celotti}, A., {Ghisellini}, G., \& {Fabian}, A.~C. 2007, \mnras, 375, 417

\bibitem[{{Cenko} {et~al.}(2006){Cenko}, {Fox}, {Moon}, {Harrison}, {Kulkarni},
  {Henning}, {Guzman}, {Bonati}, {Smith}, {Thicksten}, {Doyle}, {Petrie},
  {Gal-Yam}, {Soderberg}, {Anagnostou}, \& {Laity}}]{cenko06}
{Cenko}, S.~B., {Fox}, D.~B., {Moon}, D.-S., {et~al.} 2006, \pasp, 118, 1396

\bibitem[{{Chincarini} {et~al.}(2003){Chincarini}, {Zerbi}, {Antonelli},
  {Conconi}, {Cutispoto}, {Covino}, {D'Alessio}, {de Ugarte Postigo},
  {Molinari}, {Nicastro}, {Tosti}, {Vitali}, {Mazzoleni}, {Sciuto}, {Stefanon},
  {Jordan}, {Burderi}, {Campana}, {Danziger}, {di Paola}, {Fernandez-Soto},
  {Fiore}, {Ghisellini}, {Goldoni}, {Israel}, {Lorenzetti}, {Mc Breen},
  {Masetti}, {Messina}, {Meurs}, {Monfardini}, {Palazzi}, {Paul}, {Pian},
  {Rodono}, {Stella}, {Tagliaferri}, {Testa}, \& {Vergani}}]{chincarini03}
{Chincarini}, G., {Zerbi}, F., {Antonelli}, A., {et~al.} 2003, The Messenger,
  113, 40

\bibitem[{{Courvoisier} {et~al.}(2003){Courvoisier}, {Walter}, {Beckmann},
  {Dean}, {Dubath}, {Hudec}, {Kretschmar}, {Mereghetti}, {Montmerle},
  {Mowlavi}, {Paltani}, {Preite Martinez}, {Produit}, {Staubert}, {Strong},
  {Swings}, {Westergaard}, {White}, {Winkler}, \& {Zdziarski}}]{courvoisier03}
{Courvoisier}, T.~J.-L., {Walter}, R., {Beckmann}, V., {et~al.} 2003, \aap,
  411, L53

\bibitem[{{Dermer} \& {Schlickeiser}(1993)}]{dermer93}
{Dermer}, C.~D. \& {Schlickeiser}, R. 1993, \apj, 416, 458

\bibitem[{{Dermer} {et~al.}(1992){Dermer}, {Schlickeiser}, \&
  {Mastichiadis}}]{dermer92}
{Dermer}, C.~D., {Schlickeiser}, R., \& {Mastichiadis}, A. 1992, \aap, 256, L27

\bibitem[{{Dolcini} {et~al.}(2005){Dolcini}, {Covino}, {Treves}, {Palazzi},
  {Pian}, {Molinari}, {Chincarini}, {Zerbi}, {Rodon{\'o}}, {Testa}, {Tosti},
  {Vitali}, {Antonelli}, {Conconi}, {Cutispoto}, {Monfardini}, {Stefanon},
  {D'Avanzo}, {Danziger}, {Fernandez-Soto}, \& {Meurs}}]{dolcini05}
{Dolcini}, A., {Covino}, S., {Treves}, A., {et~al.} 2005, \aap, 443, L33

\bibitem[{{Dunlop} \& {Peacock}(1990)}]{dunlop90}
{Dunlop}, J.~S. \& {Peacock}, J.~A. 1990, \mnras, 247, 19

\bibitem[{{Elvis} {et~al.}(1989){Elvis}, {Wilkes}, \& {Lockman}}]{elvis89}
{Elvis}, M., {Wilkes}, B.~J., \& {Lockman}, F.~J. 1989, \aj, 97, 777

\bibitem[{{Fiore} {et~al.}(1998){Fiore}, {Elvis}, {Giommi}, \&
  {Padovani}}]{fiore98}
{Fiore}, F., {Elvis}, M., {Giommi}, P., \& {Padovani}, P. 1998, \apj, 492, 79

\bibitem[{{Ghisellini} {et~al.}(2002){Ghisellini}, {Celotti}, \&
  {Costamante}}]{ghisellini02}
{Ghisellini}, G., {Celotti}, A., \& {Costamante}, L. 2002, \aap, 386, 833

\bibitem[{{Ghisellini} \& {Madau}(1996)}]{ghisellini96}
{Ghisellini}, G. \& {Madau}, P. 1996, \mnras, 280, 67

\bibitem[{{Ghisellini} \& {Tavecchio}(2009)}]{ghisellini09}
{Ghisellini}, G. \& {Tavecchio}, F. 2009, ArXiv e-prints

\bibitem[{{Greiner} {et~al.}(2008){Greiner}, {Bornemann}, {Clemens}, {Deuter},
  {Hasinger}, {Honsberg}, {Huber}, {Huber}, {Krauss}, {Kr{\"u}hler},
  {K{\"u}pc{\"u} Yolda{\c s}}, {Mayer-Hasselwander}, {Mican}, {Primak},
  {Schrey}, {Steiner}, {Szokoly}, {Th{\"o}ne}, {Yolda{\c s}}, {Klose}, {Laux},
  \& {Winkler}}]{greiner08}
{Greiner}, J., {Bornemann}, W., {Clemens}, C., {et~al.} 2008, \pasp, 120, 405

\bibitem[{{Hartman} {et~al.}(1999){Hartman}, {Bertsch}, {Bloom}, {Chen},
  {Deines-Jones}, {Esposito}, {Fichtel}, {Friedlander}, {Hunter}, {McDonald},
  {Sreekumar}, {Thompson}, {Jones}, {Lin}, {Michelson}, {Nolan}, {Tompkins},
  {Kanbach}, {Mayer-Hasselwander}, {M{\"u}cke}, {Pohl}, {Reimer}, {Kniffen},
  {Schneid}, {von Montigny}, {Mukherjee}, \& {Dingus}}]{hartman99}
{Hartman}, R.~C., {Bertsch}, D.~L., {Bloom}, S.~D., {et~al.} 1999, \apjs, 123,
  79

\bibitem[{{Kalberla} {et~al.}(2005){Kalberla}, {Burton}, {Hartmann}, {Arnal},
  {Bajaja}, {Morras}, \& {P{\"o}ppel}}]{kalberla05}
{Kalberla}, P.~M.~W., {Burton}, W.~B., {Hartmann}, D., {et~al.} 2005, \aap,
  440, 775

\bibitem[{{Lebrun} {et~al.}(2003){Lebrun}, {Leray}, {Lavocat}, {Cr{\'e}tolle},
  {Arqu{\`e}s}, {Blondel}, {Bonnin}, {Bou{\`e}re}, {Cara}, {Chaleil}, {Daly},
  {Desages}, {Dzitko}, {Horeau}, {Laurent}, {Limousin}, {Mathy}, {Mauguen},
  {Meignier}, {Molini{\'e}}, {Poindron}, {Rouger}, {Sauvageon}, \&
  {Tourrette}}]{lebrun03}
{Lebrun}, F., {Leray}, J.~P., {Lavocat}, P., {et~al.} 2003, \aap, 411, L141

\bibitem[{{Lund} {et~al.}(2003){Lund}, {Budtz-J{\o}rgensen}, {Westergaard},
  {Brandt}, {Rasmussen}, {Hornstrup}, {Oxborrow}, {Chenevez}, {Jensen},
  {Laursen}, {Andersen}, {Mogensen}, {Rasmussen}, {Om{\o}}, {Pedersen},
  {Polny}, {Andersson}, {Andersson}, {K{\"a}m{\"a}r{\"a}inen}, {Vilhu},
  {Huovelin}, {Maisala}, {Morawski}, {Juchnikowski}, {Costa}, {Feroci},
  {Rubini}, {Rapisarda}, {Morelli}, {Carassiti}, {Frontera}, {Pelliciari},
  {Loffredo}, {Mart{\'{\i}}nez N{\'u}{\~n}ez}, {Reglero}, {Velasco}, {Larsson},
  {Svensson}, {Zdziarski}, {Castro-Tirado}, {Attina}, {Goria}, {Giulianelli},
  {Cordero}, {Rezazad}, {Schmidt}, {Carli}, {Gomez}, {Jensen}, {Sarri},
  {Tiemon}, {Orr}, {Much}, {Kretschmar}, \& {Schnopper}}]{lund03}
{Lund}, N., {Budtz-J{\o}rgensen}, C., {Westergaard}, N.~J., {et~al.} 2003,
  \aap, 411, L231

\bibitem[{{Mannheim}(1993)}]{mannheim93}
{Mannheim}, K. 1993, \aap, 269, 67

\bibitem[{{M{\"u}cke} {et~al.}(2003){M{\"u}cke}, {Protheroe}, {Engel},
  {Rachen}, \& {Stanev}}]{muecke03}
{M{\"u}cke}, A., {Protheroe}, R.~J., {Engel}, R., {Rachen}, J.~P., \& {Stanev},
  T. 2003, Astroparticle Physics, 18, 593

\bibitem[{{O'Flaherty} \& {Jakobsen}(1997)}]{oflaherty97}
{O'Flaherty}, K.~S. \& {Jakobsen}, P. 1997, \apj, 479, 673

\bibitem[{{Rees}(1966)}]{rees66}
{Rees}, M.~J. 1966, \nat, 211, 468

\bibitem[{{Reeves} \& {Turner}(2000)}]{reeves00}
{Reeves}, J.~N. \& {Turner}, M.~J.~L. 2000, \mnras, 316, 234

\bibitem[{{Reeves} {et~al.}(2001){Reeves}, {Turner}, {Bennie}, {Pounds},
  {Short}, {O'Brien}, {Boller}, {Kuster}, \& {Tiengo}}]{reeves01}
{Reeves}, J.~N., {Turner}, M.~J.~L., {Bennie}, P.~J., {et~al.} 2001, \aap, 365,
  L116

\bibitem[{{Roming} {et~al.}(2005){Roming}, {Kennedy}, {Mason}, {Nousek}, {Ahr},
  {Bingham}, {Broos}, {Carter}, {Hancock}, {Huckle}, {Hunsberger}, {Kawakami},
  {Killough}, {Koch}, {McLelland}, {Smith}, {Smith}, {Soto}, {Boyd},
  {Breeveld}, {Holland}, {Ivanushkina}, {Pryzby}, {Still}, \&
  {Stock}}]{roming05}
{Roming}, P.~W.~A., {Kennedy}, T.~E., {Mason}, K.~O., {et~al.} 2005, Space
  Science Reviews, 120, 95

\bibitem[{{Sambruna} {et~al.}(2007){Sambruna}, {Tavecchio}, {Ghisellini},
  {Donato}, {Holland}, {Markwardt}, {Tueller}, \& {Mushotzky}}]{sambruna07}
{Sambruna}, R.~M., {Tavecchio}, F., {Ghisellini}, G., {et~al.} 2007, \apj, 669,
  884

\bibitem[{{Schartel} {et~al.}(1997){Schartel}, {Komossa}, {Brinkmann}, {Fink},
  {Truemper}, \& {Wamsteker}}]{schartel97}
{Schartel}, N., {Komossa}, S., {Brinkmann}, W., {et~al.} 1997, \aap, 320, 421

\bibitem[{{Schlegel} {et~al.}(1998){Schlegel}, {Finkbeiner}, \&
  {Davis}}]{schlegel98}
{Schlegel}, D.~J., {Finkbeiner}, D.~P., \& {Davis}, M. 1998, \apj, 500, 525

\bibitem[{{Siebert} {et~al.}(1996){Siebert}, {Matsuoka}, {Brinkmann}, {Cappi},
  {Mihara}, \& {Takahashi}}]{siebert96}
{Siebert}, J., {Matsuoka}, M., {Brinkmann}, W., {et~al.} 1996, \aap, 307, 8

\bibitem[{{Sikora} {et~al.}(1994){Sikora}, {Begelman}, \& {Rees}}]{sikora94}
{Sikora}, M., {Begelman}, M.~C., \& {Rees}, M.~J. 1994, \apj, 421, 153

\bibitem[{{Sowards-Emmerd} {et~al.}(2004){Sowards-Emmerd}, {Romani},
  {Michelson}, \& {Ulvestad}}]{sowards-emmerd04}
{Sowards-Emmerd}, D., {Romani}, R.~W., {Michelson}, P.~F., \& {Ulvestad}, J.~S.
  2004, \apj, 609, 564

\bibitem[{{Tavecchio} {et~al.}(2007){Tavecchio}, {Maraschi}, {Ghisellini},
  {Kataoka}, {Foschini}, {Sambruna}, \& {Tagliaferri}}]{tavecchio07}
{Tavecchio}, F., {Maraschi}, L., {Ghisellini}, G., {et~al.} 2007, \apj, 665,
  980

\bibitem[{{Tody}(1993)}]{tody93}
{Tody}, D. 1993, in Astronomical Society of the Pacific Conference Series,
  Vol.~52, Astronomical Data Analysis Software and Systems II, ed. R.~J.
  {Hanisch}, R.~J.~V. {Brissenden}, \& J.~{Barnes}, 173--+

\bibitem[{{Ulrich} {et~al.}(1997){Ulrich}, {Maraschi}, \& {Urry}}]{ulrich97}
{Ulrich}, M.-H., {Maraschi}, L., \& {Urry}, C.~M. 1997, \araa, 35, 445

\bibitem[{{Urry} \& {Padovani}(1995)}]{urry95}
{Urry}, C.~M. \& {Padovani}, P. 1995, \pasp, 107, 803

\bibitem[{{Wall} \& {Jackson}(1997)}]{wall97}
{Wall}, J.~V. \& {Jackson}, C.~A. 1997, \mnras, 290, L17

\bibitem[{{Weymann} {et~al.}(1981){Weymann}, {Carswell}, \&
  {Smith}}]{weymann81}
{Weymann}, R.~J., {Carswell}, R.~F., \& {Smith}, M.~G. 1981, \araa, 19, 41

\bibitem[{{Willott} {et~al.}(2001){Willott}, {Rawlings}, {Blundell}, {Lacy}, \&
  {Eales}}]{willott01}
{Willott}, C.~J., {Rawlings}, S., {Blundell}, K.~M., {Lacy}, M., \& {Eales},
  S.~A. 2001, \mnras, 322, 536

\bibitem[{{Worsley} {et~al.}(2004{\natexlab{a}}){Worsley}, {Fabian}, {Celotti},
  \& {Iwasawa}}]{worsley04a}
{Worsley}, M.~A., {Fabian}, A.~C., {Celotti}, A., \& {Iwasawa}, K.
  2004{\natexlab{a}}, \mnras, 350, L67

\bibitem[{{Worsley} {et~al.}(2004{\natexlab{b}}){Worsley}, {Fabian}, {Turner},
  {Celotti}, \& {Iwasawa}}]{worsley04b}
{Worsley}, M.~A., {Fabian}, A.~C., {Turner}, A.~K., {Celotti}, A., \&
  {Iwasawa}, K. 2004{\natexlab{b}}, \mnras, 350, 207

\bibitem[{{Wright} {et~al.}(1978){Wright}, {Peterson}, {Jauncey}, \&
  {Condon}}]{wright78}
{Wright}, A.~E., {Peterson}, B.~A., {Jauncey}, D.~L., \& {Condon}, J.~J. 1978,
  \apjl, 226, L61

\bibitem[{{Yuan} {et~al.}(2005){Yuan}, {Fabian}, {Celotti}, {McMahon}, \&
  {Matsuoka}}]{yuan05}
{Yuan}, W., {Fabian}, A.~C., {Celotti}, A., {McMahon}, R.~G., \& {Matsuoka}, M.
  2005, \mnras, 358, 432

\bibitem[{{Zamorani} {et~al.}(1981){Zamorani}, {Henry}, {Maccacaro},
  {Tananbaum}, {Soltan}, {Avni}, {Liebert}, {Stocke}, {Strittmatter},
  {Weymann}, {Smith}, \& {Condon}}]{zamorani81}
{Zamorani}, G., {Henry}, J.~P., {Maccacaro}, T., {et~al.} 1981, \apj, 245, 357

\end{thebibliography}

\end{document}